\documentclass[aps,prl,twocolumn,amsmath,amssymb,floatfix]{revtex4}
\usepackage{tabularx}
\usepackage{bm}
\usepackage{euscript}
\usepackage{epsfig,psfrag,subfigure}
\usepackage{graphicx}
\usepackage{color}
\usepackage{amsfonts}
\usepackage{exscale}
\usepackage{amsbsy}
\usepackage{dsfont}

\begin{document}
\title{Fe-based superconductors: unity or diversity?}
\author{Steven A. Kivelson and Hong Yao}
\affiliation{Department of Physics, Stanford University, Stanford, California 94305}

\maketitle

{\bf Does the high temperature superconductivity observed in the newly discovered iron-pnictide materials represent another example of the same essential physics responsible for superconductivity in the cuprates, or does it embody a new mechanism?}

\begin{figure}[!ht]
\subfigure[]{
\includegraphics[scale=0.30]{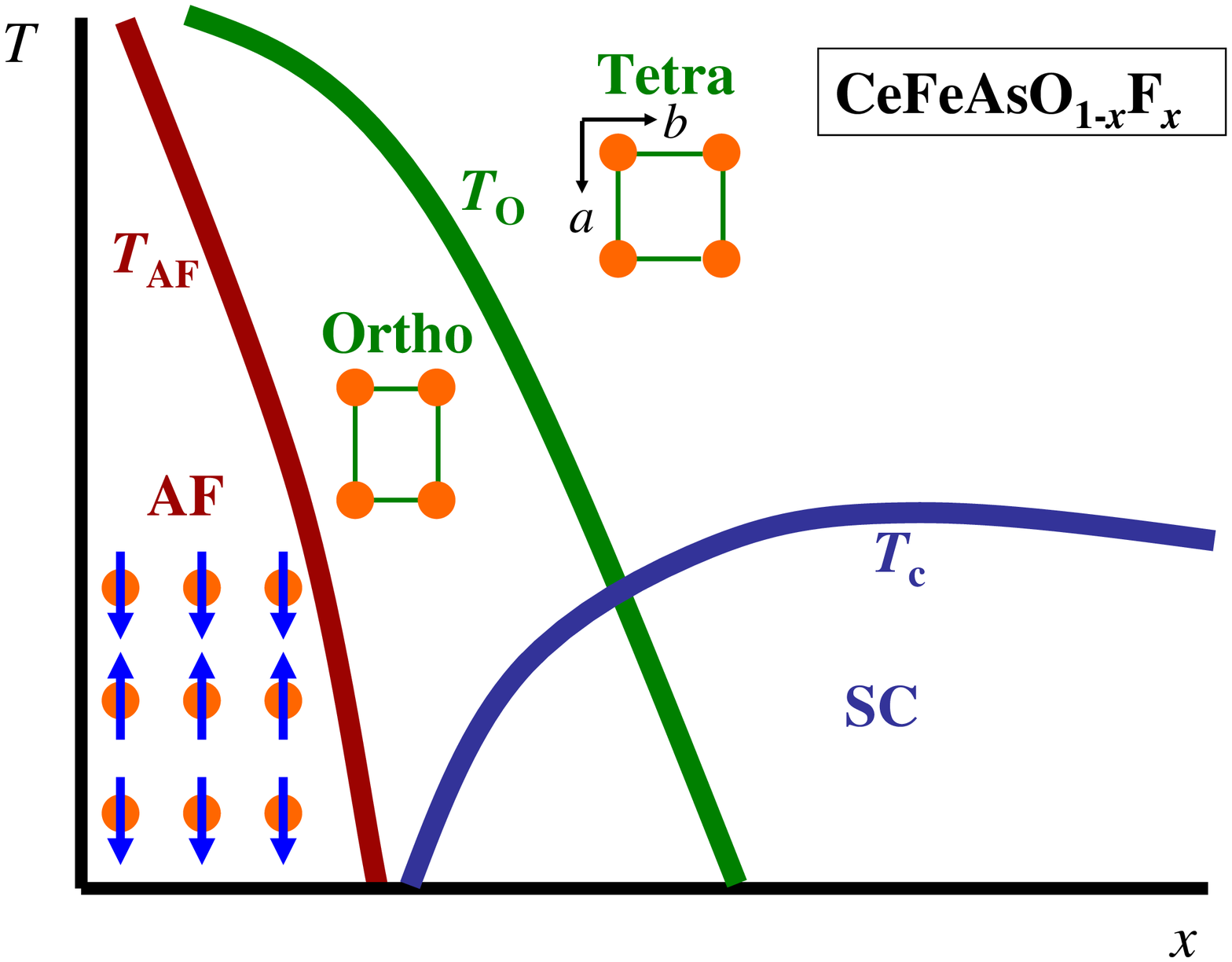}}
\subfigure[]{
\includegraphics[scale=0.30]{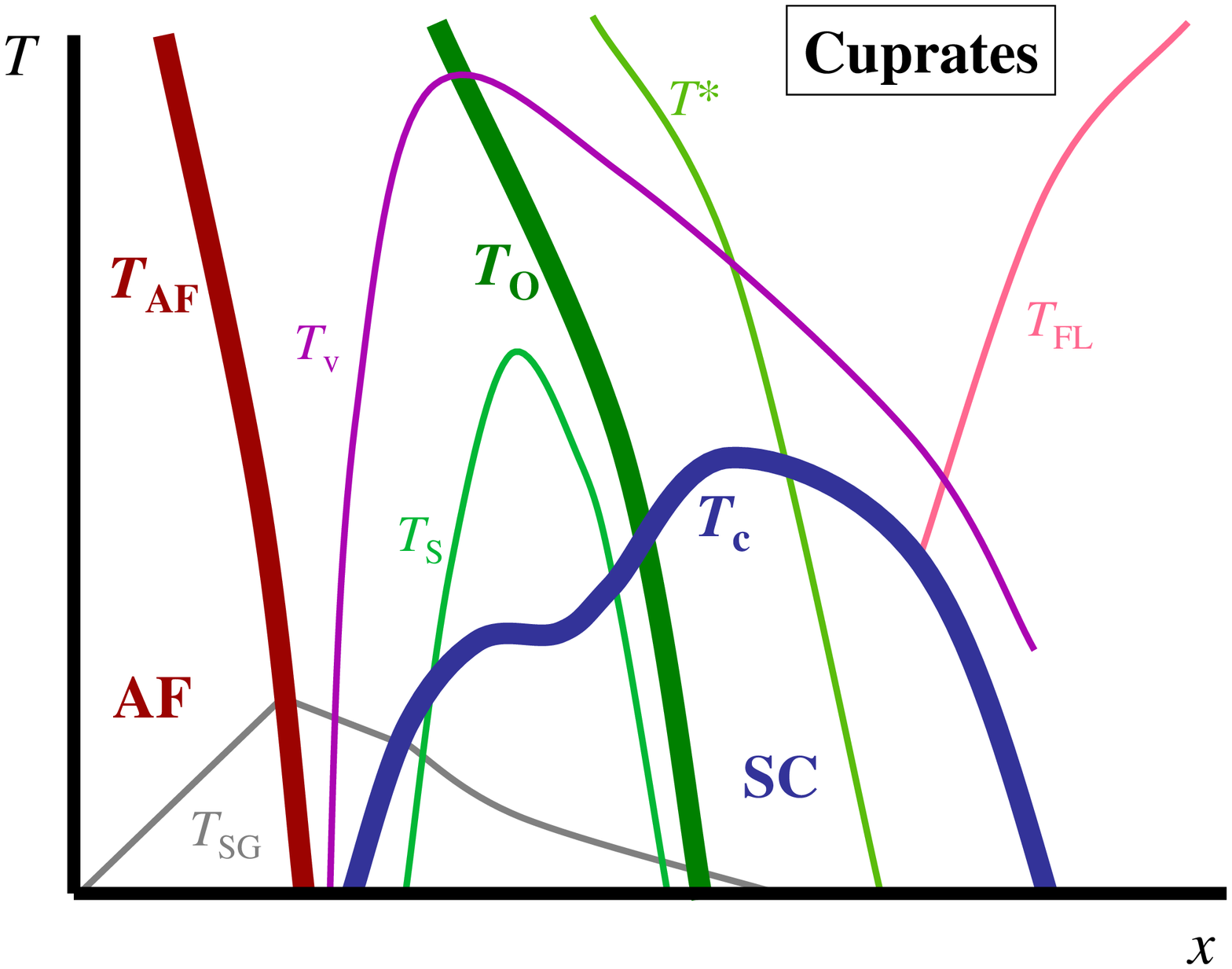}}
\caption{Phase diagrams of the iron pnictide and cuprate high temperature superconductors: (a) Schematic phase diagram of CeFeAsO$_{1-x}$F$_x$ as a function of doped hole concentration, $x$, and temperature, $T$. The temperature $T_{\textrm{AF}}$ delineates the boundary of
the antiferromagnetically ordered phase, $T_{\textrm{O}}$ indicates the boundary of the structural distorted (orthorhombic or ``nematic'') phase, and $T_\textrm{c}$ is the superconducting transition temperature. (b) Schematic composite phase diagram
of the cuprate high-temperature superconductors as a function of doped hole concentration, $x$, and temperature,
$T$. The temperatures $T_{\textrm{AF}}$, $T_{\textrm{O}}$ and $T_{\textrm{c}}$ are as in Fig. 1a. Other crossover or transition lines indicated on the figure are: $T_{\textrm{SG}}$, spin-glass transition; $T_{\textrm{S}}$, stripe ordering transition; $T_{\textrm{v}}$, vortex liquid/Nernst effect crossover; $T^*$, pseudo-gap crossover or transition; $T_{\textrm{FL}}$, Fermi liquid crossover. In La$_{2-x}$Sr$_x$CuO, $T_{\textrm{O}}$ represents a structural phase transition to a
low-temperature orthorhombic phase, whereas in YBCO it represents the nematic crossover found by Hinkov {\it et al.} \cite{Hinkov08}.}
\end{figure}

Superconductivity is among the most fascinating properties a material can exhibit. On the fundamental level, it represents a direct, macroscopic manifestation of coherent quantum mechanical behavior, while its potential practical importance is almost unlimited, especially if new superconductors can be synthesized or discovered with still higher transition temperatures. For the past two decades, the search for high temperature superconductivity was narrowly focused on the study of a family of cuprate perovskites, relatives of the original superconducting cuprate discovered in 1986 by Bednorz and Mueller.  That changed with the discovery \cite{Kamihawa08}, this year, of a new family of iron-pnictide high temperature superconductors with $T_\textrm{c}$ as high as 55K.  There have already been hundreds of papers written on the subject, a testament to the excitement generated by this discovery.

In drawing broader inferences concerning mechanisms of high temperature superconductivity, a key question is whether the essential physics of the iron pnictides is the same or different as that operating in the cuprates. In a paper in this issue, Zhao {\it et al.} \cite{Zhao08} report the results of a systematic neutron scattering study of the structural and magnetic phase diagram of CeFeAsO$_{1-x}$F$_x$ as a function of ``doping concentration,'' $x$, and temperature, $T$.  The result is shown schematically in Fig. 1a. This result is significant in its own right. In addition to superconductivity, two other distinct forms of broken symmetry (illustrated in the insets of the figure) are known to exist in the iron-pnictides:  antiferromagnetic order involves a pattern of alternating magnetic moments (``up and down spins'');  orthorhombic order is associated with a structural distortion such that the lattice constant in the $a$-direction is different than that in the $b$-direction. Using experiments (neutron scattering) which directly measure these types of order, Zhao {\it et al.} have studied samples over a range of $x$ so as to map out the full phase diagram.

A significant conclusion drawn by Zhao {\it et al.} is that there are striking similarities between the iron-pnictide and the cuprate phase diagrams.  By inference, then, it is suggested that there are important similarities in the essential physics, including the mechanism of superconductivity. Clearly, this is a conclusion with potentially broad consequences. However, there have already been many arguments made, both in favor and against this analogy in other papers.  A trouble with such arguments is that it is not, a priori, clear which aspects of the problem are key, so any specific difference or similarity is (legitimately) given different weight by different authors.  At least the present paper allows us to add to the list of similarities.

In both materials, there is an ``undoped'' parent compound that is antiferromagnetically ordered.  In the cuprartes, the antiferromagnet is strongly insulating, while in the iron-pnictides it is metallic, although with an unreasonably low conductivity for a metal.  In both materials, the antiferromagnetic order is collinear, and doubles the size of the unit cell, albeit with slightly different magnetic structure. One interesting new observation is that the magnitude of the ordered moment is always substantial, yet it varies considerably among the various iron-pnictides; the moment in the iron-pnictides ranges from about 0.5 to 1.5 times its value in the cuprates. Moreover, in both materials, as a function of increasing doping, the antiferromagnetic critical temperature, $T_{\textrm{AF}}$, drops precipitously, and vanishes near the point at which superconductivity onsets.

The structural transition temperature, $T_{\textrm{O}}$ in the iron-pnictides is initially very slightly larger than $T_{\textrm{AF}}$, and then decreases as the doping is increased, although less precipitously than $T_{\textrm{AF}}$.  Indeed, Zhao {\it et al.} have found that $T_\textrm{O}$ vanishes at around the same value of $x$ that the superconducting critical temperature, $T_{\textrm{c}}$, reaches its maximum value, which is suggestive of an intimate connection between the two types of order. (Similar behavior is also observed in the systematic doping study of another iron-pnictide LaFeAsO$_{1-x}$F$_x$ \cite{Huang08}, which suggests that this behavior may be generic in iron-pnictides.) A similar structural phase diagram is seen in at least some materials in the cuprate family (e.g. La$_{2-x}$Sr$_x$CuO$_4$.).  The most interesting comparison, which was not made by Zhao {\it et al.}, is to measurements on the underdoped cuprate superconductor YBCO, which imply the existence of an electronic nematic phase \cite{Hinkov08} with a critical temperature which vanishes, seemingly at a quantum critical point, somewhere under the superconducting dome, as illustrated by the thick green line in Fig. 1b. [A nematic phase, like the phase below $T_{\textrm{O}}$ in the iron-pnictides, is one that spontaneously breaks the square symmetry of the underlying ideal crystal, resulting in a rectangular (or related) symmetry in which the $a$ and $b$ axes are inequivalent.]

The correspondence between the roles of the structurally distorted phase in CeAsFeO$_{1-x}$F$_x$ and the nematic phase in YBCO is one of the potentially most significant aspects \cite{Fang08,Xu08} of the Zhao {\it et al.} paper. However, partly because various cuprates have different crystal structures to begin with, and partly because there have been so many studies of these materials, the number of transition and crossover lines that have been identified in the phase diagram of the cuprates is enormous, as is illustrated, only slightly tongue in cheek, in Fig. 1b.  If one looks hard enough, one can find in the cuprates something that is reminiscent of almost any interesting phenomenon in solid state physics.

The paper of Zhao {\it et al.} offers no new information concerning the internal symmetries of the superconducting state. Some experiments, which find evidence of residual gapless ``nodal'') quasiparticle excitations at low temperatures, suggest that it has $d$-wave symmetry (as does the superconducting state in the cuprates), and some suggest that it has a full gap, suggestive of a more conventional $s$-wave symmetry. Intriguingly, it has recently been suggested \cite{Mazin08,JP08,Wang08} on theoretical grounds that the superconducting order has an ``unconventional'' $s$-wave symmetry, with an order parameter that changes sign in going from one Fermi surface pocket to another.  At the most basic level, the pairing symmetry may be less significant than whether or not the order parameter changes sign as a function of momentum. Any unconventional order is plausibly related to strong local repulsions between electrons. In contrast, a conventional $s$-wave order most likely indicates that the pairing arises from induced attractions between electrons, as in conventional low temperature superconductors. Clearly, the strength of the analogy between the two families of high temperature superconductor will be much clearer when the symmetries of the superconducting state in the iron-pnictides have been established.

\end{document}